\documentclass[12pt,preprint]{aastex}

\begin{document}

\title{Astrometric Constraints on the Masses of Long-Period Gas Giant 
Planets in the TRAPPIST-1 Planetary System}

\author{Alan P.~Boss$^1$, Alycia J.~Weinberger$^1$, Sandra A.~Keiser$^{1\dagger}$,
Tri L. Astraatmadja$^1$, Guillem Anglada-Escude$^2$, and Ian B.~Thompson$^3$}

\affil{$^1$Department of Terrestrial Magnetism, Carnegie Institution for
Science, 5241 Broad Branch Road, NW, Washington, DC 20015-1305, USA}

\affil{$^2$School of Physics and Astronomy, Queen Mary University of London,
327 Mile End Road, London, E1 4NS, UK}

\affil{$^3$Carnegie Observatories, Carnegie Institution for Science,
813 Santa Barbara Street, Pasadena, CA 91101-1292, USA}

\email{aboss@carnegiescience.edu}

\begin{abstract}

 Transit photometry of the M8V dwarf star TRAPPIST-1 
(2MASS J23062928-0502285) has revealed
the presence of at least seven planets with masses and radii similar
to that of Earth orbiting at distances that might allow liquid water
to be present on their surfaces. We have been following TRAPPIST-1
since 2011 with the CAPSCam astrometric camera on the 2.5-m du Pont
telescope at the Las Campanas Observatory in Chile. In 2016 we noted
that TRAPPIST-1 lies slightly farther away than previously thought, at 
12.49 pc, rather than 12.1 pc. Here we examine fifteen epochs of CAPSCam 
observations of TRAPPIST-1, spanning the five years from 2011 to 2016, 
and obtain a revised trigonometric distance of $12.56 \pm 0.12$ pc.
The astrometric data analysis pipeline shows no evidence for a long-period
astrometric wobble of TRAPPIST-1. After proper motion and parallax
are removed, residuals at the level of $\pm 1.3$ millarcsec (mas)
remain. The amplitude of these residuals constrains the masses of any
long-period gas giant planets in the TRAPPIST-1 system: no planet
more massive than $\sim 4.6 M_{Jup}$ orbits with a 1 yr period, and
no planet more massive than $\sim 1.6 M_{Jup}$ orbits with a 5 yr period.
Further refinement of the CAPSCam data analysis pipeline, combined
with continued CAPSCam observations, should either detect any 
long-period planets, or put an even tighter constraint on these mass 
upper limits.

\end{abstract}

\keywords{astrometry — planetary systems — stars: low mass}

\section{Introduction}

 Since 2007 we have been engaged in a long-term program using the Carnegie 
Astrometric Planet Search Camera (CAPSCam) on the Las Campanas 
Observatory 2.5-m du Pont telescope to search for gas giant planets in 
orbit around nearby low-mass stars by the astrometric detection method.
There are 21 known G stars within 10 pc of the sun and at least 236 M 
dwarfs (Henry et al. 1997), with nearby late M dwarfs continuing to be 
discovered (e.g., Reiners \& Basri 2009). These abundant M dwarfs are a 
natural choice for astrometric planet searches, because low mass primaries 
and close proximity to the sun greatly enhance the detectability of planetary
companions.

 M dwarfs have become the favored targets for finding the closest habitable 
worlds, those most amenable to follow-up observations. M dwarf exoplanets
are likely to be the primary targets to be searched by JWST, the James 
Webb Space Telescope, for atmospheric biosignatures (Deming et al. 2009) 
among the transiting super-Earths to be found by TESS, the Transiting 
Exoplanet Survey Satellite (Ricker et al. 2014). In fact, M dwarfs are 
estimated to harbor about 37\% of the habitable zone
``real estate'' within 10 pc of the Sun (Cantrell et al. 2013), and
estimates of the frequency of habitable Earths around M dwarfs
can be as high as 53\% (Kopparapu 2013). The detection of a 
habitable-zone exoEarth orbiting the M6V dwarf Proxima Centuari 
(Anglada-Escud\'e et al. 2016), literally the closest star to the sun
at 1.295 pc, has raised the stakes in the race for the first direct
imaging and spectroscopy studies of nearby habitable worlds.

 The detection of at first three (Gillon et al. 2016), and then a total of
at least seven exoplanets (Gillon et al. 2017) orbiting the M8V dwarf
TRAPPIST-1 (2M2306-05, 2MASS J23062928-0502285)
has only added fuel to this race to discover and characterize
potentially habitable exoplanets. Transit photometry has revealed
that five of the seven known TRAPPIST-1 planets have radii and masses within
factors of $\sim 1.1$ and $\sim 2$, respectively, of that of the Earth,
and three of these appear to have stellar irradiation levels that might
permit the existence of oceans of surface water, assuming Earth-like atmospheres
(Gillon et al. 2017). 

 The near-resonant orbital periods of the six innermost TRAPPIST-1 
planets suggests their formation at a greater orbital distance followed
by inward coupled migration (Gillon et al. 2017). A number of the first 
hot and warm super-Earth exoplanets discovered by Doppler spectroscopy 
were found to have at least one longer period gas giant sibling planet, 
with several having two (e.g., the M4V dwarf Gl 876 and the G8V star
HD 69830) or even three such siblings (e.g., the G3V star Mu Ara and
the G8V star 55 Cnc). HD 181433 is a K5V star with an inner 7.5 Earth-mass 
planet and two outer Jupiter-mass planets, while the G6V HD 47186 has 
an inner 22 Earth-mass planet and an outer Saturn-mass planet 
(Bouchy et al. 2009). Outer gas giants may thus accompany inner habitable 
worlds, a situation analogous to that of our own solar system. 

 The question then becomes, what other planets might be orbiting 
TRAPPIST-1 at greater distances than the seven known planets? The
outermost one, TRAPPIST-1 h, has a semi-major axis of $\sim 0.063$ AU,
leaving a large amount of orbital distance unexplored.
Montet et al. (2014) combined Doppler and direct imaging results to 
estimate that about 6.5\% of M dwarfs host one or more gas giants within 
20 AU. Gas giants orbiting M dwarfs may represent a challenge to the core 
accretion formation mechanism for gas giant planet formation 
(e.g., Koshimoto et al. 2014), but not for the competing disk 
instability mechanism (e.g., Boss 2006). The VLT FORS2 camera astrometric 
survey of Sahlmann et al. (2014) set an upper bound of about 9\% on 
the occurrence of gas giants larger than 5 Jupiter masses orbiting 
at 0.01 to 0.8 AU around a sample of 20 M8 to L2 dwarfs. The {\it Gaia} space
telescope is presently searching about 1000 M dwarfs for long-period exoplanets 
(Sozzetti et al. 2014; Perryman et al. 2014).

 We began observing TRAPPIST-1 with CAPSCam
in 2011 as part of a long-term astrometric program to detect gas giant 
planets around approximately 100 nearby M, L, and T dwarf stars. 
Weinberger et al. (2016) published a trigonometric distance for TRAPPIST-1
of 12.49 pc $\pm 0.2$ pc, slightly larger than the previous trigonometric
distance of 12.1 pc $\pm 0.4$ pc (Costa et al. 2006), but consistent
within the error bars. Reiners and Basri (2009) found a spectrophotometric 
distance of 12.1 pc. The CAPSCam distance was based on the first eleven epochs
of observations, spanning the 3 yrs from 2011 to 2014. Here we
re-analyze TRAPPIST-1, using the additional 4 epochs of observations
taken in 2015 and 2016 to refine the trigonometric parallax and proper
motions derived by Weinberger et al. (2016) and to search for any evidence
of long-period gas giant companions to TRAPPIST-1.

\section{Astrometric Camera}

 The CAPSCam camera (Boss et al. 2009) employs a Hawaii-2RG HyViSi hybrid 
array (2048 x 2048) that allows the definition of an arbitrary Guide Window 
(GW), which can be read out (and reset) rapidly, repeatedly, and 
independently of the rest of the array, the Full Frame (FF). The GW is 
centered on the relatively bright target stars, with multiple short exposures to
avoid saturation. The rest of the array then integrates for prolonged
periods on the background reference grid of fainter stars.
The natural plate scale for CAPSCam on the 2.5-m du Pont is 0.196 
arcsec/pixel, a scale that allows us to avoid introducing any
extra optical elements into the system that would produce astrometric
errors. An astrometric quality ($\lambda/30$) dewar filter window with a 
red bandpass of 810 nm to 910 nm (similar to SDSS z)
is the only other optical element in
the system besides the du Pont primary and secondary mirrors.
We take multiple exposures with CAPSCam, with small variations (2 arcsec) of 
the image position (dithering to four positions), in order to average out uncertainties due to pixel response non-uniformity. 

 Analysis of the star NLTT 48256 showed that an astrometric accuracy better 
than 0.4 milliarcsec (mas) can be obtained over a time scale of several years 
with CAPSCam on the du Pont 2.5-m (Boss et al. 2009). Anglada-Escud\'e et al. 
(2012) used CAPSCam to place an upper mass limit on the known Doppler 
exoplanet GJ 317b (Johnson et al. 2007). Analysis of the M3.5 dwarf 
GJ 317 showed that the limiting astrometric accuracy, at least for the 
brighter targets in our sample, is about 0.6 mas per epoch in Right Ascension 
(R.A.) and about 1 mas per epoch in Declination (Dec.), based on the
results of the preliminary data analysis pipeline available at the time.
With 18 epochs, the overall accuracy in fitting the parallax of
GJ 317 was about 0.15 mas (Anglada-Escud\'e et al. 2012).
For comparison, Very Large Telescope (VLT) astrometry with SPHERE yields 
positional errors of about 1 mas (Zurlo et al. 2014). Hence CAPSCam
should be able to either detect or place significant astrometric
constraints on the masses of any long-period gas giants in the TRAPPIST-1
system.

\section{Observations and Data Reduction}

 TRAPPIST-1 (RA = 23 06 29.283; DEC = -05 02 28.59 [2000]) was observed
at the 15 epochs listed in Table 1 between 2011 and 2016. TRAPPIST-1 is
bright enough (I= 14.024, J= 11.354) that the
GW mode was used, with either a 10 sec or 15 sec GW exposure (depending
on the seeing conditions: the former for seeing of $\sim 1.0$ arcsec,
and the latter for seeing of $\sim 1.4$ arcsec) and a 90 sec FF exposure.
At each epoch, there were typically 20 frames taken during a 40 minute
time period as TRAPPIST-1 passed through the meridian.
The data were analyzed using the same data pipeline that has been used
for reducing the data for all previously published CAPSCam observations,
the APTa pipeline developed by one of us (GAE). Details about ATPa may
be found in Boss et al. (2009) and in Anglada-Escud\'e et al. (2012) and 
about the parallax, proper motions, and astrometric zero point corrections 
in Weinberger et al. (2016).

 Weinberger et al. (2016) used the same first eleven epochs of
observations as are used here to determine an absolute parallax 
$\pi_{abs} = 80.09 \pm 1.17$ mas and proper motions in R.A. and Dec.,
respectively, of $922.02 \pm 0.61$ mas yr$^{-1}$ and $-461.88 \pm 0.94$ 
mas yr$^{-1}$. This absolute parallax is based on a relative
parallax of $\pi_{rel} = 79.10 \pm 1.11$ mas and a zero point
correction of $-0.99 \pm 0.36$ mas. Note that the proper motions 
are not corrected for any bias in the reference frame, and that
Weinberger et al. (2016) used only a portion of the ATPa data analysis 
pipeline and developed new routines to fit for the parallax and proper
motion. When all fifteen epochs listed in Table 1 are included in 
the solution, the absolute parallax becomes 
$\pi_{abs} = 79.59 \pm 0.78$ mas and the proper motions in R.A. and Dec.,
respectively, become $922.64$ mas yr$^{-1}$ and $-462.88$ mas yr$^{-1}$.
Now the absolute parallax is based on a relative
parallax of $\pi_{rel} = 78.51 \pm 0.75$ mas and a zero point
correction of $-1.08 \pm 0.22$ mas, as described below. 
Clearly the revised values for the parallax and proper motion are 
consistent with each other and are well within the error bars.
TRAPPIST-1 thus has a revised trigonometric parallax distance of
$12.56 \pm 0.12$ pc.

 The correction to absolute parallax was based on a comparison of the photometric distance to five reference stars with good catalog photometry 
at V, I, J, H, and Ks and with fit effective temperatures $\ge$ 4000 K.  
Our CAPSCam measurements of these stars yield nominal parallaxes, which 
in this case are all smaller than their uncertainties. We averaged the 
offset between the nominal parallaxes and the photometric parallaxes 
obtained by fitting Kurucz stellar atmosphere models to the broad-band 
photometry and find a correction to absolute parallax of $1.08 \pm 0.22$ 
mas. Note that this correction is irrelevant to the determination of the 
residuals to the five parameter fit to the astrometry and therefore 
does not contribute to the upper limit on any giant planet companion 
to the star.

\section{Astrometric Residuals}

 Figure 1 displays the CAPSCam TRAPPIST-1 field image that served as
the reference plate for the ATPa astrometric analysis. TRAPPIST-1 itself
lies within the central GW, while the stars labelled with blue numbers
were used as the reference stars for the ATPa analysis. Red vectors
depict the inferred proper motions of the stars.
 
 Table 1 lists the ATPa residuals in R.A. and Dec. for the apparent motion
of TRAPPIST-1 with respect to the background reference stars that remain
after the parallax and proper motion have been removed from the solution.
Table 1 also lists the individual epoch statistical uncertainties, 
computed as the root mean square (RMS) of the offsets of the individual 
frames that are taken at each epoch. We note that the standard deviation of the 
residuals ($\sim$ 1.2 mas in R.A.) is much larger than the typical individual 
epoch statistical uncertainty ($\sim$ 0.6 mas), and we attribute the difference 
to systematic sources that we are working to resolve.

 The ATPa pipeline searches for periodicities in these Table 1 residuals 
that could be caused by a companion object on a circular orbit, with a 
minimum orbital period of 80 days and maximum orbital period of 4000 days.
Figure 2 shows the resulting power spectrum for a possible long-period
companion to TRAPPIST-1. The power spectrum yields no hints of any
long-period planets, as it is dominated by a large number of short-period
peaks with rather low amplitudes, with little power at periods longer
than about 1500 days. 

 Figures 3 and 4 plot the ATPa residuals ($\delta_{R.A.}, \delta_{Dec.}$)
from Table 1, allowing a visual search for any suspected periodicity.
These figures confirm what is evident from Figure 2, namely that the
only way to fit the residuals with an astrometric wobble would be to
use an orbit with a period of order a year or less. Given that the
fifteen CAPSCam observations spanning a little over 5 years average out to
only about 3 epochs per year, orbital periods of a year or less are
likely to spurious. Ongoing work on developing the CAPSCam date pipeline
will allow us to further refine the astrometric analysis.

 In Table 1, the mean absolute value of the ATPa residuals in R.A. is 
1.17 mas, while that for the residuals in Dec. is 1.46 mas. For the
detection of an astrometric amplitude $A$ (in mas) in the presence of
astrometric uncertainties of $\sigma$ (in mas) with $N_{obs}$ observations,
a conservative estimate of the signal to noise ratio ($SNR$) is given by

$$ SNR = { A \over \sigma } \sqrt{N_{obs}}. $$

\noindent
Taking the larger of the two residuals, namely those in Dec., as
$\sigma = 1.46$ mas, $N_{obs} = 15$, and requiring $SNR = 5$ yields
a conservative estimate of the largest amplitude that could be hidden
in our CAPSCam data of $A = 1.9$ mas.

 We then use 1.9 mas as an upper limit on any wobble of TRAPPIST-1, allowing 
us to place upper limits on the mass of any long-period gas giant companions,
as a function of their orbital periods. The angular semimajor axis ($\Theta$) 
of the displacement of a star about the common center of mass of 
a star-planet system is given (e.g., Boss 1996) by 

$$ \Theta = { M_p a \over M_s r} = { M_p P^{2/3} \over M_s^{2/3} r}, $$

\noindent
where $\Theta$ is in arcsec, $M_p$ is the planet mass (in solar masses), 
$M_s$ is the stellar mass (in solar masses), $M_p$ is assumed to
be negligible in mass compared to $M_s$, $a$ is the orbital semimajor axis
(in AU) and is equal to the radius of an assumed circular orbit, $P$ is 
the orbital period (in years), and $r$ is the distance to the 
star-planet system (in pc). For example, with only Jupiter orbiting
around the sun, the solar wobble is $\pm$ 1 mas when viewed from 5 pc.
Note that in the case of TRAPPIST-1, where the primary has a mass of 
only about 80 $M_{Jup}$ (Gillon et al. 2017), neglecting the planet mass 
is not strictly a good approximation when considering possible
planetary masses in the multiple Jupiter-mass range, but this approximation
is adequate for the constraint considered here.

 Figure 5 displays the possible long-period planet masses that appear to
be ruled out by our non-detection at the 1.9 mas level for the TRAPPIST-1
system, compared to the six TRAPPIST-1 planets with known masses, and all the
confirmed exoplanets currently contained in the NASA Exoplanet Archive. 
We assumed a mass for TRAPPIST-1 of 0.08 $M_\odot$ (Gillon et al. 2017). 
We halt the astrometric constraint at an orbital period of five years,
the length of our CAPSCam observations.
It can be seen from Figure 5 that our astrometric constraint still leaves a 
large area of discovery space to be explored for additional planets in 
the TRAPPIST-1 system. While the plethora of confirmed exoplanets
shown in Figure 5 is suggestive of the possibility of other planets to be
found around TRAPPIST-1, it should be noted that most of those confirmed planets
orbit stars of earlier type than the M8 dwarf TRAPPIST-1, and planetary
demographics can be expected to depend on stellar masses.

\section{Doppler Constraints}

 Tanner et al. (2012) performed Keck II NIRSPEC Doppler spectroscopy of
23 late-M dwarfs, including 2M2306-05 = TRAPPIST-1. Their data consisted
of three epochs, one in 2006 and two more in 2010 taken
on consecutive nights, a sequence chosen to search for both short-period
hot Jupiters and long-period cold Jupiters. They obtained  
radial velocities for 2M2306-05 with an
uncertainty of 130 m s$^{-1}$. Following Boss (1996), the radial
velocity amplitude in km s$^{-1}$ for an edge-on planetary orbit is 

$$ v_r = 30 { M_p \over P^{1/3} M_s^{2/3}}, $$

\noindent
using the same units as previously. Figure 5 shows the resulting
upper limit on the mass of any undetected compansions to TRAPPIST-1,
based on the conservative estimate of a maximum Doppler wobble of 
380 m s$^{-1}$. This estimate assumes that $SNR = 5$ (Mayor \& Queloz 
1995), $\sigma = 130$ m s$^{-1}$, and $N_{obs} = 3$. It can be
seen that this Doppler upper mass limit, in combination with our 
astrometric upper mass limit, and rules out a large portion of possible
discovery space, though an equally large and interesting portion
remains to be explored. Orbital stability of the TRAPPIST-1 system
appears to be assured even in the presence of a 15 $M_\oplus$ planet,
provided that it orbits beyond 0.37 AU (Quarles et al. 2017), yielding
another constraint on any undiscovered planets in this system.

 We have also performed a more involved analysis of the companion masses 
ruled out by the Tanner et al. (2012) Doppler RV observations. To assess 
the detectability of RV companions, we followed the procedure described in 
Kohn et al. (2016). We calculated orbits for 482 million putative binary 
systems for masses of the secondary of 0.6 to 12.6 $M_{Jup}$ and periods 
up to 1550 days. The mass of the primary was assumed to be 0.08 $M_\odot$. 
For each period and mass-ratio pair, we calculated the radial velocities that 
would be observed given a single epoch precision of 130 m s$^{-1}$
for an ensemble of binaries with orbital elements drawn from the 
eccentricity distribution for known planets as described by the Beta distribution 
in Kipping et al. (2013), a uniform distribution of $sin \ i$ (inclination $i$), 
a uniform distribution of time of periastron passage over the period, 
and a uniform distribution of the longitude of periastron. We define 
a binary to be observable at a given period and mass-ratio if, 
in 67\% of the calculated orbits, the RMS of its calculated RV 
exceeded 225 m s$^{-1}$. We plot the resulting RV constraint as a blue 
curve in Figure 5 as well.

\section{Photometry}

 Given that about two transits occur every day in the TRAPPIST-1 system
(Gillon et al. 2017), and that the transits last for at least a half-hour 
apiece, transits should be occurring on average for one hour every day.
Each CAPSCam epoch lasted for roughly one hour on target, so 15 epochs
observed meant that CAPSCam had observed TRAPPIST-1 for a total of
about 15 hours, or 0.6 day. Hence there is a reasonable chance that at
least one transit occurred during our astrometric observations.

 We thus decided to check to see if any transits were obvious in our 
CAPSCam data. The deepest transit depth for a TRAPPIST-1 planet
is about 0.782\% (for TRAPPIST-1 g), while the shallowest depth is
about 0.352\% (for TRAPPIST-1 h), so detecting a transit requires
millimag photometry from the ground. CAPSCam was not designed to be
a photometric camera, and our astrometric observations do not require
photometric seeing, and as a consequence our photometric precision
has not been measured or developed. A quick look at a total of 356 
CAPSCam images of TRAPPIST-1 found that when the photometric flux
in TRAPPIST-1 was normalized by that of the brighter reference stars
in the 6.63 arcmin x 6.63 arcmin field of view, variations of order
unity occurred in an irregular manner. Even larger variations were
evident when normalized to fainter reference stars. As result, we
abandoned our search for transits in the CAPSCam data.

\section{Conclusions}

 The TRAPPIST-1 system is a fascinating example of a planetary system
that will continue be a focus of intensive research in the coming decades.
Our ongoing astrometric observations with CAPSCam have refined
the trigonometric distance to TRAPPIST-1, with the new value being
12.56 pc. In addition, the absence of any clear, long-period signals
in the residuals, once the parallax and proper motions have been
removed, places strong upper limits on the masses of any gas giant
planets orbiting well beyond the seven known transiting planets:
no planets more massive than $\sim 4.6 M_{Jup}$ orbit with a 1 yr period, 
and none more massive than $\sim 1.6 M_{Jup}$ orbit with a 5 yr period.
A large region of discovery space intermediate between these long-period
orbits and the short-period orbits of the TRAPPIST-1 planets, however, 
remains to be explored by other means.

 One of us (TLA) is presently working on further developing ATPa, with the
goal of reducing sources of systematic errors, such as those caused
by differential chromatic refraction, and by distortions of the entire 
optical system that might change with time. The latter is being addressed 
by taking multiply dithered (4 x 4) CAPSCam images of rich stellar fields 
at multiple epochs, and generating a pixel by pixel correction function. 
We plan to continue our astrometric observations of TRAPPIST-1, and
we look forward to learning what an improved analysis of our data might
reveal about the TRAPPIST-1 planetary system.

\acknowledgments

 We thank the David W. Thompson Family Fund for support of the
CAPSCam astrometric planet search program, the Carnegie Observatories
for continued access to the du Pont telescope, and the telescope
operators and technicians at the Las Campanas Observatory for making
these observations possible. We also thank the referee for several 
suggestions for improvements. The development of the CAPSCam camera was
supported in part by NSF grant AST-0352912.

\clearpage
\begin{deluxetable}{cccccc}
\tablecaption{Log of epochs in Julian days (J2000) and calendar
days (YYYYMMDD format) of CAPSCam observations 
of TRAPPIST-1 as analyzed by ATPa and the residuals ($\delta_{R.A.}, 
\delta_{Dec.}$) in R.A. and Dec., respectively, after removing the parallax
and proper motion, and the uncertainties ($\pm \delta_{R.A.}, 
\pm \delta_{Dec.}$) in those residuals, all in milliarcsec (mas).
\label{tbl-1}}
\tablehead{\colhead{\quad Julian Date \quad} & 
\colhead{\quad \quad Epoch \quad \quad} &
\colhead{\quad $\delta_{R.A.}$ \quad \quad} &
\colhead{\quad $\pm \delta_{R.A.}$ \quad \quad} &
\colhead{\quad $\delta_{Dec.}$ \quad \quad} &
\colhead{\quad $\pm \delta_{Dec.}$ \quad \quad} }
\startdata

2455787.79 & 20110814  &  1.61    &  0.503  &   -2.75   &  0.542\\

2455841.67 & 20111007  &  -1.64   &  0.522  &   -0.923  &  0.903  \\

2455842.64 & 20111008  &  -1.18   &  0.603  &    0.538  &  0.574 \\

2456085.94 & 20120607  &  0.280   &  0.483  &    0.353  &  0.645 \\

2456134.87 & 20120726  & -1.79    &  1.49   &    5.23   &  1.83  \\

2456194.65 & 20120924  &  2.68    &  0.573  &   -0.673  &  0.708 \\

2456489.84 & 20130715  & -0.410   &  0.378  &    0.0471 &  0.802   \\

2456519.75 & 20130814  &  -0.110  &  0.217  &   -0.540  &  0.357 \\

2456552.69 & 20130916  &  -0.030  &  0.824  &    3.16   &  1.08   \\

2456586.57 & 20131021  & -1.70    &  0.462  &    2.76   &  0.857 \\

2456888.75 & 20140819  & 1.15     &  0.472  &   -2.51   &  0.537 \\

2457232.80 & 20150729  & -0.624   &  0.678  &   -0.126  &  0.738 \\

2457283.66 & 20150918  & 1.08     &  0.572  &    0.601  &  0.601 \\

2457562.87 & 20160623  & -1.51    &  0.347  &    0.801  &  0.789 \\

2457667.62 & 20161006  & -1.77    &  0.593  &   -0.897  &  0.729 \\

\enddata
\end{deluxetable}
\clearpage

\begin{figure}
\vspace{-2.0in}
\includegraphics[scale=.60,angle=0]{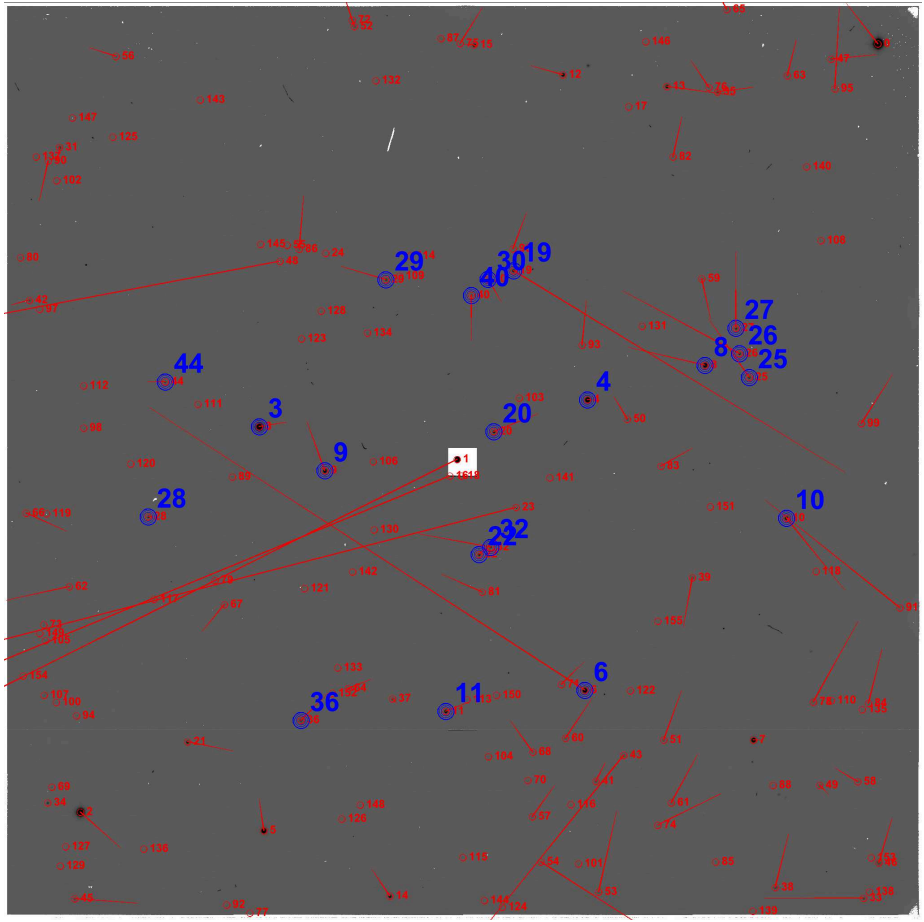}
\caption{CAPSCam image of TRAPPIST-1 field used as the reference plate for 
the ATPa data pipeline analysis. TRAPPIST-1 is labelled as star 1 and
can be seen inside the central Guide Window (GW). Red vectors denote inferred
proper motions from the ATPa analysis. The 20 blue numbered stars were used as
the reference stars for the analysis, while the stars labelled only with
red numbers were not. Star 19 is a typical reference star with a brightness
roughly equal to the median, with I = 18.1 and J = 15.8.}
\end{figure}

\clearpage

\begin{figure}
\vspace{-2.0in}
\includegraphics[scale=.60,angle=0]{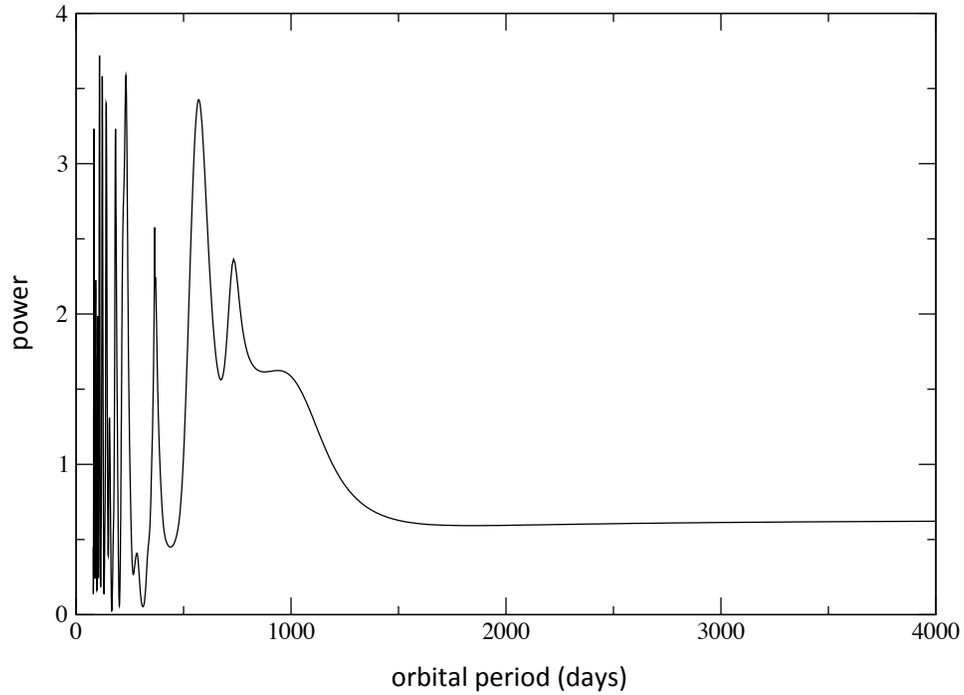}
\caption{Power spectrum of the ATPa residuals for TRAPPIST-1 as a function of assumed
secondary companion orbital period in days. The fact that there are multiple 
peaks with relatively low power at short orbital periods during the $\sim$ 
2000-day span of CAPSCam observations places an upper limit on the masses
of any gas giant companions with orbital periods of $\sim$ 6 yr or less.}
\end{figure}

\clearpage

\begin{figure}
\vspace{-2.0in}
\includegraphics[scale=.60,angle=0]{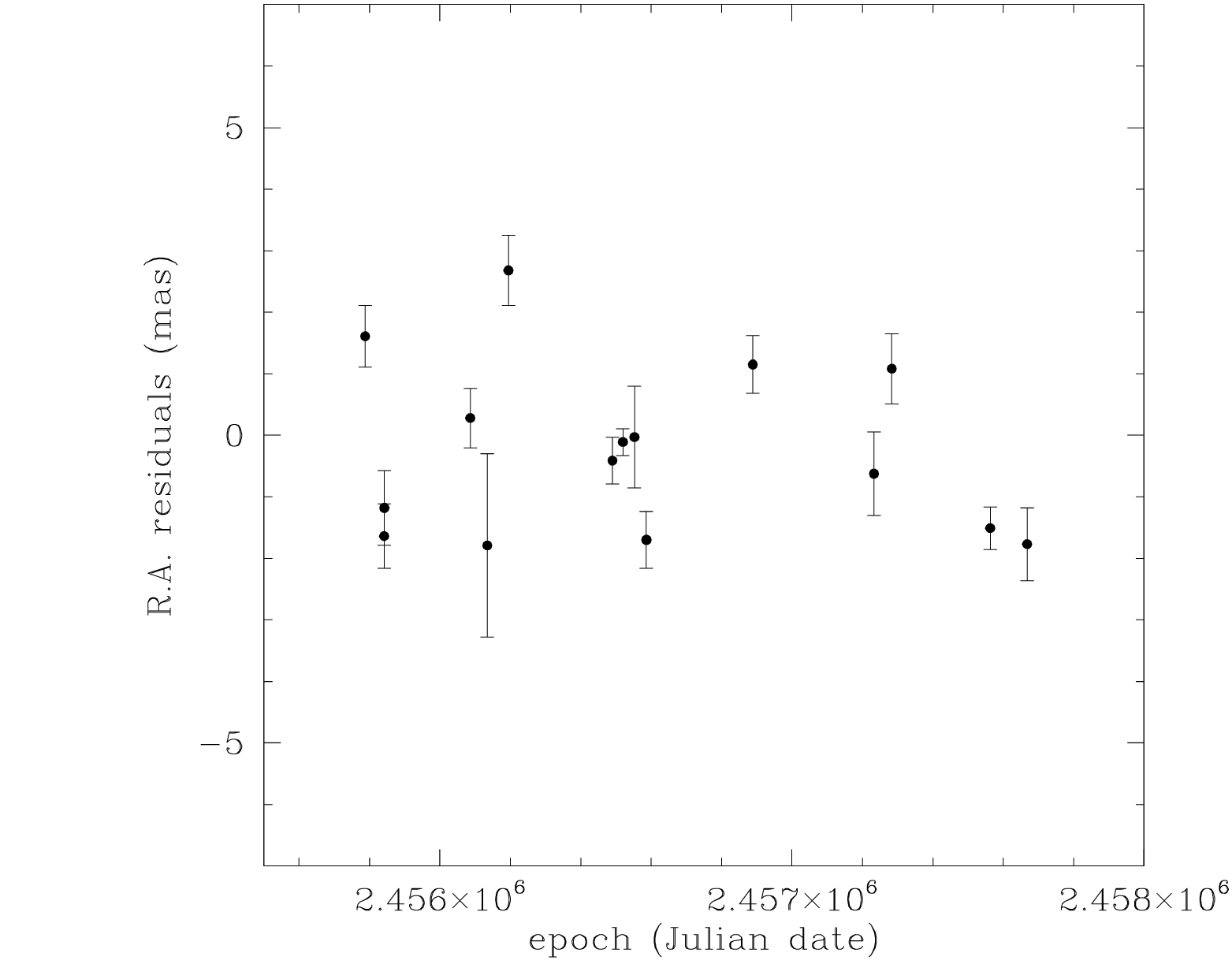}
\caption{Astrometric ATPa residuals for TRAPPIST-1 in R.A. as a function of epoch in Julian days, spanning a period of over five years of CAPSCam observations.}
\end{figure}

\clearpage

\begin{figure}
\vspace{-2.0in}
\includegraphics[scale=.60,angle=0]{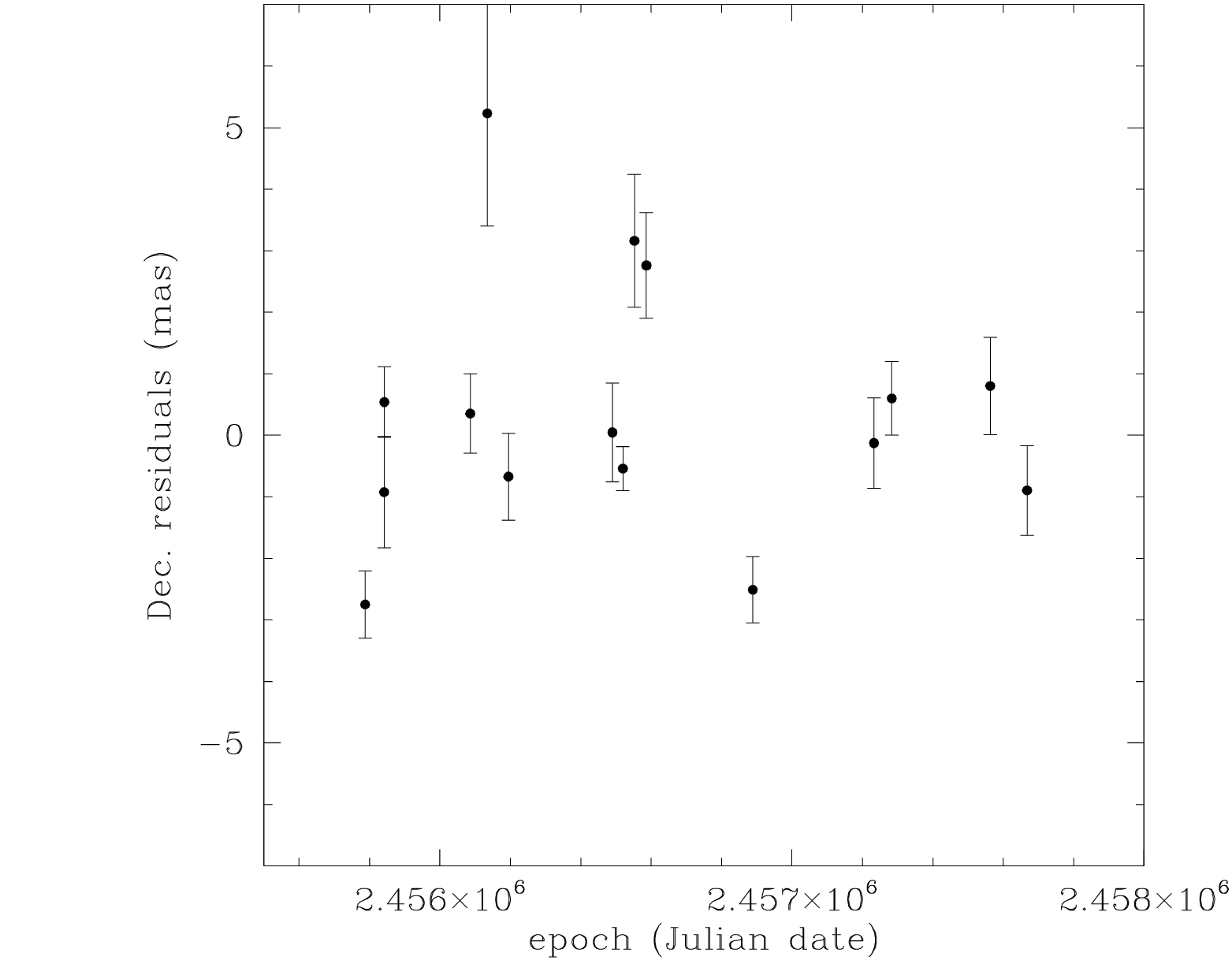}
\caption{Astrometric ATPa residuals for TRAPPIST-1 in Dec. as a function of epoch in Julian days, spanning a period of over five years of CAPSCam observations.}
\end{figure}

\clearpage

\begin{figure}
\vspace{-2.0in}
\includegraphics[scale=.60,angle=0]{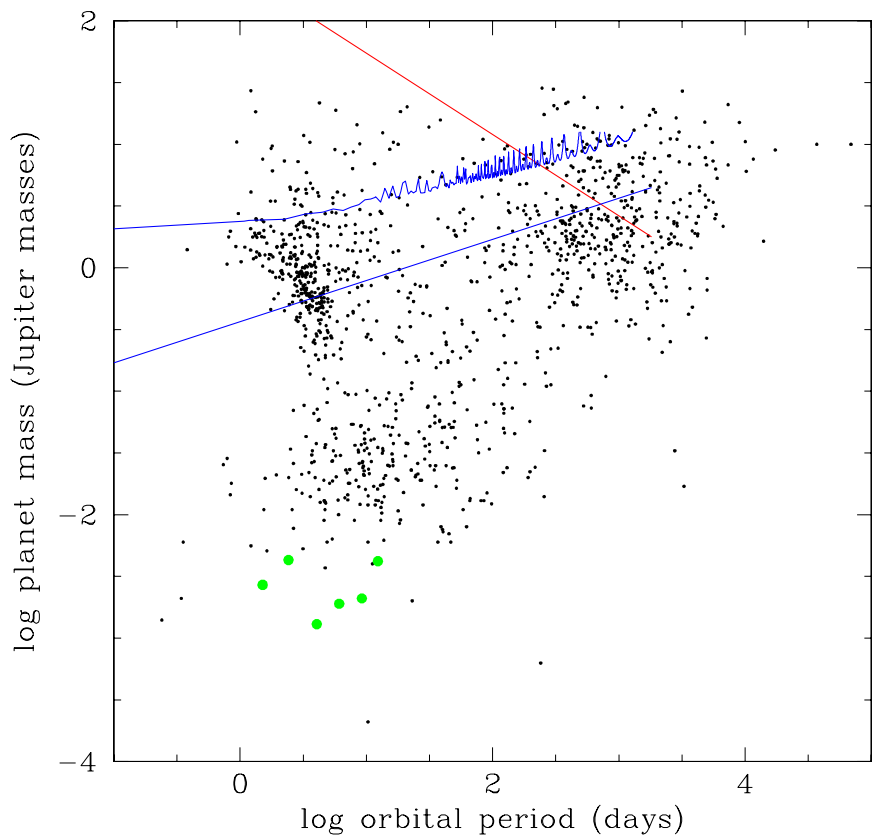}
\caption{The region above the red line is the portion of discovery 
space for long-period TRAPPIST-1 planets that appears to be eliminated 
by our CAPSCam observations, conservatively assuming a maximum astrometric 
wobble of $\pm$ 1.9 mas and periods less than 5 years,
compared to the six TRAPPIST-1 planets with known masses (green dots) 
and all of the confirmed exoplanets known as of May 25, 2017, 
from the NASA Exoplanet Archive database (black dots).
The lower blue line is the simple upper mass limit constraint placed 
by Doppler spectroscopy (Tanner et al. 2012), corresponding to a 
radial velocity variation of $\pm$ 380 m s$^{-1}$, while the upper blue
curve is the limit derived from the more detailed analysis.}
\end{figure}

\end{document}